\documentclass[12pt]{iopart}
\usepackage{graphicx}

\newcommand{\ignore}[1]{}
\begin{document}

\title[Magnetic softness in iron-based superconductors]{Magnetic softness in iron-based superconductors}

\author{Wei-Guo Yin*, Chi-Cheng Lee, and Wei Ku}

\address{Condensed Matter Physics and Materials Science Department, Brookhaven National Laboratory, Upton, New York 11973, USA}
\ead{wyin@bnl.gov}
\begin{abstract}
We examine the relevance of several major material-dependent parameters to the magnetic softness in iron-base superconductors by first-principles electronic structure analysis of their parent compounds. The results are explained in the spin-fermion model where localized spins and orbitally degenerate itinerant electrons coexist and are coupled by Hund's rule coupling. We found that the difference in the strength of the Hund's rule coupling term is the major material-dependent microscopic parameter for determining the ground-state spin pattern. The magnetic softness in iron-based superconductors is essentially driven by the competition between the double-exchange ferromagnetism and the superexchange antiferromagnetism.
\end{abstract}

\maketitle

\section{Introduction}

\begin{figure}[b]
\center{
\includegraphics[width=0.6\columnwidth,clip=true,angle=0]{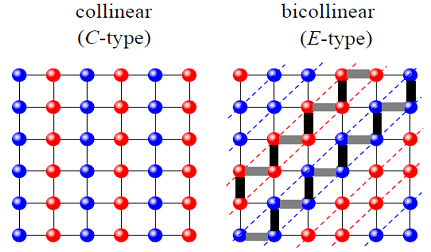}
}
\caption{\label{fig1} Schematics of two AF spin orders in the iron plane. Red and blue balls stand for spin-up and spin-down Fe atoms, respectively. Dashed lines represent the `bicollinear' view. The zigzag horizontal and vertical solid thick lines represent the $E$-type view \cite{_DE:11:Yin} following the context of manganites \cite{Mn:Hotta}.}
\end{figure}

Recently, high transition-temperature ($T_c$) superconductivity has been observed in a number of doped iron-based layer materials \cite{syn:1111:Kamihara,syn:122:Rotter,syn:11:Yeh,syn:se122:Guo} near a static antiferromagnetic (AF) order \cite{neu:1111:Cruz,neu:122:Huang,neu:11:Li,neu:11:Bao,neu:K0.8Fe1.6Se2:Bao} and with a spin resonance \cite{neu:122:Christianson,neu:11:Qiu,neu:se122:Park}, a pattern exhibited previously by the layered cuprate high-$T_c$ superconductors. It has been generally believed that strong spin fluctuations in two-dimensional (2D) space is at the heart of the high-$T_c$ mechanism. A proper understanding of the magnetism in parent undoped materials thus becomes an essential first step towards understanding the high-$T_c$ mechanism. The undoped cuprates are universally described by the 2D Heisenberg model. The key character of this model is that a novel resonating-valence-bond spin-liquid state competes fiercely with the traditional N\'{e}el spin-solid state for being the ground state \cite{Cu:Anderson}, a situation referred to as magnetic softness. Magnetic softness becomes even more apparent in the parent compounds of iron-based superconductors (FeSCs), since different ground-state AF spin patterns are truly realized---`collinear' $C$-type in iron pnictides (e.g., LaOFeAs and BaFe$_2$As$_2$ \cite{neu:1111:Cruz,neu:122:Huang}) and `bicollinear' $E$-type in iron chalcogenides (e.g., FeTe$_{1-x}$Se$_x$ \cite{neu:11:Li,neu:11:Bao}), as illustrated in Fig.~\ref{fig1}---despite apparent similarity in crystal and electronic structures. To elucidate its nature, it is necessary to first identify what material-dependent parameter drives the difference in the magnetic pattern; this knowledge will put stringent constraints on minimum theoretical modeling of FeSCs.

In literature, the anion height from the iron plane ($z_\mathrm{anion}$) and the ordered magnetic moment ($m$) have been considered as major material-dependent parameters \cite{Tc:Mizuguchi,_strong:dmft:Yin_NM}, since neutron diffraction experiments reported $z_\mathrm{anion}=$
1.31, 1.35, and 1.73 {\AA} and $m=$ 0.36, 0.87, and 1.70 $\mu_B$ in
LaOFeAs, BaFe$_2$As$_2$, and FeTe, respectively \cite{neu:1111:Cruz,neu:122:Huang,neu:11:Li,neu:11:Bao}. First-principles studies qualitatively reproduced these observations \cite{dft:1111:Yildirim,dft:11:Ma} and further revealed that the $C$-$E$ magnetic transition can be induced by tuning $z_\mathrm{anion}$ in FeTe$_{1-x}$Se$_x$ \cite{dft:11:Moon}. Since varying $z_\mathrm{anion}$ also varies $m$, it is unclear whether it is $m$ or something else that determines the $C$-$E$ transition. Another interesting scenario is that the difference in the orbital ordering pattern determines the ground-state magnetic pattern: one of the initially degenerate Fe $d_{xz}$ and $d_{yz}$ orbitals gets more populated than the other in the $C$-type pnictides \cite{_oo:Kruger,dft:1111:Lee,_oo:Lv,_oo:Chen}, and Fe $d_{YZ}=(d_{xz}-d_{yz})/\sqrt{2}$ orbital was stipulated to be more populated than $d_{XZ}=(d_{xz}+d_{yz})/\sqrt{2}$ in the $E$-type FeTe \cite{_oo:Turner:11}. While the former was verified by first-principles calculations \cite{dft:1111:Lee}, the latter remains untested.

The purpose of this paper is to examine the relevance of $z_\mathrm{anion}$, $m$, and orbital order to the magnetic softness in the parent compounds of FeSCs by first-principles electronic structure analysis. We show that increasing $z_\mathrm{anion}$ in BaFe$_2$As$_2$ also increases $m$ and drives the magnetic transition from $C$ to $E$, but that varying $m$ and fixing $z_\mathrm{anion}$ cannot. We further demonstrate that the $E$ type does not exhibit a site orbital order. These results are explained by use of a recently proposed spin-fermion model \cite{_DE:11:Yin} where localized spins and orbitally degenerate itinerant electrons coexist and are coupled by Hund's rule coupling \cite{_DE:1111:Kou,_DE:Lv,so:1111:Wu,_strong:Si}. We found that the difference in the strength of the Hund's rule coupling term is the major material-dependent microscopic parameter for determining the ground-state spin pattern in the parent compounds of FeSCs. The effort to gain the Hund's rule coupling energy is known to induce the double-exchange ferromagnetism \cite{Mn:Zener,Mn:Anderson}. Therefore, the magnetic softness in FeSCs is essentially driven by the competition between the double-exchange ferromagnetism and the superexchange antiferromagnetism in the localized-spins sector.

\begin{figure}[t]
\center{
\includegraphics[width=0.7\columnwidth,clip=true,angle=0]{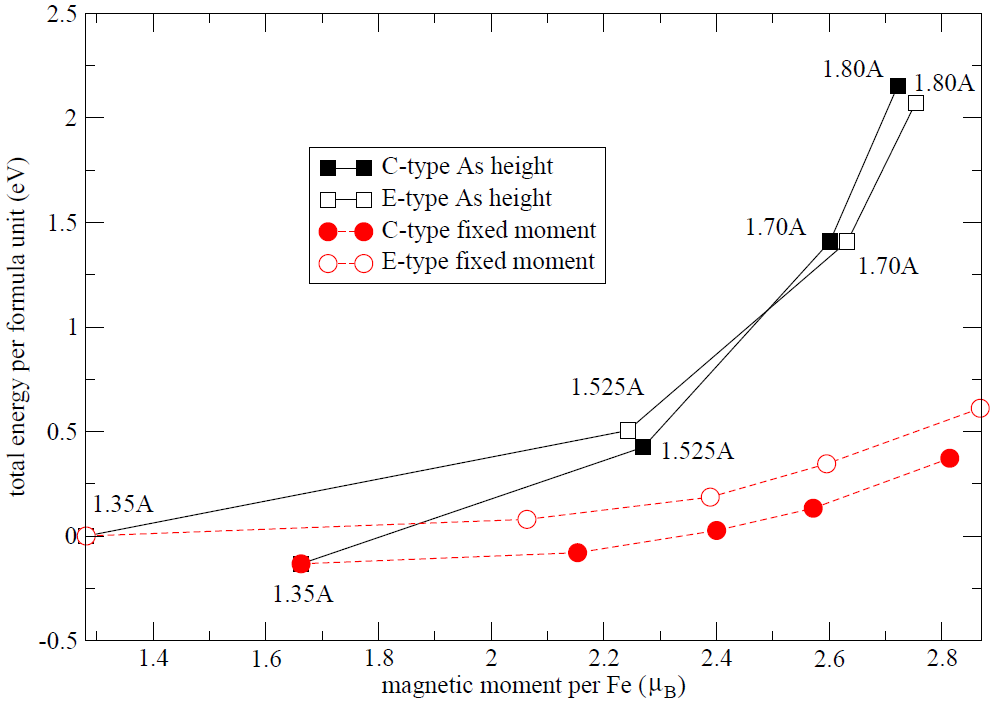}
}
\caption{\label{fig2} Comparing the total energies of the $C$-type and $E$-type AF states in BaFe$_2$As$_2$ as a function of $m$ obtained by fixed-spin-moment calculations (circles) and by varying $z_\mathrm{anion}$ (squares). In the latter case, $z_\mathrm{anion}$ is marked near the corresponding datum.}
\end{figure}

\section{First-principles analysis}

We performed first-principles calculations in local spin-density approximation of density functional theory with full potential, all-electron, linearized augmented plane wave basis implemented in the WIEN2k software package \cite{wien2k}. We adopted an eight-Fe-atom unit cell for all the calculations and the $10\times 10 \times 8$ mesh of k points. The energy convergence is better than $1$ meV per Fe atom.

In Fig.~\ref{fig2}, we compare the total energies of the $C$-type and $E$-type AF states in BaFe$_2$As$_2$. First, the effect of $z_\mathrm{anion}$ was investigated. Like in FeTe$_{1-x}$Se$_x$ \cite{dft:11:Moon}, increasing $z_\mathrm{anion}$ in BaFe$_2$As$_2$ also increases $m$ and drives the magnetic transition from $C$ to $E$ around $z_\mathrm{anion}=1.6$ {\AA}.  Then, to isolate the effect of $m$ from $z_\mathrm{anion}$, we fixed $z_\mathrm{anion}$ at the experimental position of 1.35 {\AA} \cite{syn:122:Rotter} and varied $m$ using the fixed-spin-moment method. We found that varying $m$ alone cannot drive the $C$-$E$ transition.

To examine a possible orbital order in the $E$ type, we performed first-principles Wannier function analysis \cite{dft:1111:Lee,Mn:lamno3:Yin} on FeTe and got the following Fe density matrix:
$$
\left(
  \begin{array}{cccccc}
         & 3z^2-r^2 & x^2-y^2 & yz & xz & xy \\
3z^2-r^2 & 1.44  & 0.00  & -0.04 &	-0.04  & 0.03   \\
x^2-y^2 &  0.00  & 1.31  & 0.04  & -0.04  & 0.00   \\
yz &      -0.04 & 0.04  & \mathbf{1.05}  & 0.01  & 0.05    \\
xz &      -0.04 & -0.04 &	0.01  & \mathbf{1.05}  & 0.05    \\
xy &       0.03  & 0.00  & 0.05  & 0.05  & 1.13    \\
  \end{array}
\right),
$$
where the $x$ and $y$ axes point to the nearest-neighbor (NN) Fe atoms. To compare with the orbital ordering pattern proposed in Ref. \cite{_oo:Turner:11}, the coordinate system needs to be rotated by $45^\circ$ in the Fe plane. The resulting density matrix is rewritten as
 $$
\left(
  \begin{array}{cccccc}
         & 3z^2-r^2 & XY & YZ & XZ & X^2-Y^2 \\
3z^2-r^2 & 1.44  & 0.00  & 0.00 &	-0.057  & 0.03   \\
XY &  0.00  & 1.31  & -0.057  & 0.00  & 0.00   \\
YZ &      0.00 & -0.057  & \mathbf{1.04}  & 0.00  & 0.00    \\
XZ &      -0.057 & 0.00 &	0.00  & \mathbf{1.06}  & 0.071    \\
X^2-Y^2 &       0.03  & 0.00  & 0.00  & 0.071  & 1.13    \\
  \end{array}
\right),
$$
where the $X$ and $Y$ axes point to the next-nearest-neighbor (NNN) Fe atoms. Apparently, orbital polarization within the $XZ$ and $YZ$ orbitals is negligible. Therefore, site orbital ordering is not the driving force for the $E$-type spin order in FeTe.

Below we explain these results using the spin-fermion model that was shown to be capable of providing a unified picture for magnetic correlation and electronic transport in the parent compounds of FeSCs \cite{_DE:11:Yin,_DE:se122:Yin,_DE:1111:Liang}.

\begin{figure}[t]
\center{
\includegraphics[width=0.6\columnwidth,clip=true,angle=0]{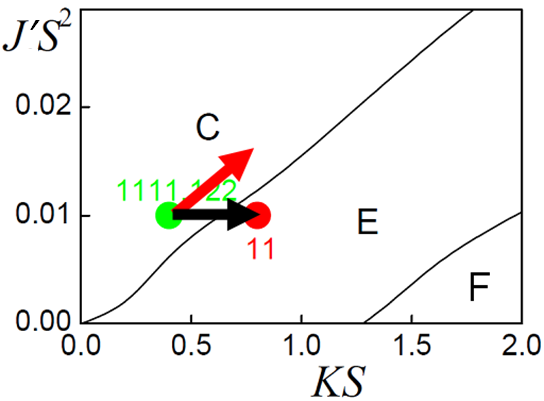}
}
\caption{\label{fig3} Phase diagram of the spin-fermion model for $J'=J$. Black and red arrows correspond to changing $z_\mathrm{anion}$ and $m$ in Fig.~\ref{fig2}, respectively. The dots marked with 1111, 122, and 11 are schematic locations of LaOFeAs, BaFe$_2$As$_2$, and FeTe, respectively.}
\end{figure}

\section{The spin-fermion model}

The electrons in the degenerate Fe $3d_{xz}$ and $3d_{yz}$ orbitals are treated as itinerant electrons, and those in the rest Fe $3d$ orbitals as effective localized spins. This leads to an effective two-orbital double-exchange model \cite{_DE:11:Yin,_DE:1111:Kou,_DE:Lv}:
\begin{eqnarray}
\label{eq1} H=&-&\sum\limits_{ij\gamma \gamma^\prime \mu }
{(t_{ij}^{\gamma \gamma^\prime} C_{i\gamma \mu }^\dag C_{j
\gamma^\prime \mu}^{} +h.c.)} \nonumber \\
&-& \frac{K}{2}\sum\limits_{i\gamma \mu \mu' } {C_{i\gamma \mu
}^\dag \vec {\sigma }_{\mu \mu' } C_{i\gamma \mu' }^{} }  \cdot \vec
{S}_i + \sum\limits_{ij} { J_{ij} \vec {S}_i \cdot \vec {S}_j},
\end{eqnarray}
where $C_{i\gamma \mu }^{} $ denotes the annihilation operator of an
itinerant electron with spin $\mu=\uparrow$ or $\downarrow $ in the
$\gamma=d_{xz}$ or $d_{yz} $ orbital on site $i$. $t_{ij}^{\gamma
\gamma^\prime} $'s are the electron hopping parameters. $\vec
{\sigma }_{\mu \mu' } $ is the Pauli matrix and $\vec {S}_i$ is the
localized spin whose magnitude is $S$. $K$ is the effective Hund's rule coupling. $J_{ij}$ is the AF
superexchange couplings between the localized spins; in particular,
$J$ and $J'$ are respectively the NN and NNN ones. The filling of the itinerant electrons is on average three (one hole) per Fe site,  corresponding to the high-spin configuration of Fe $3d^6$ \cite{dft:1111:Lee}. To the $y$ direction, the $d_{xz}$-$d_{xz}$ NN hopping integral $t_\| \simeq 0.4$ eV and the $d_{yz}$-$d_{yz}$ NN hopping integral $t_\bot \simeq 0.13$ eV; they are swapped to the $x$ direction; by symmetry the NN interorbital hoppings are zero; the NNN intraorbital hopping integral $t'\simeq -0.25$ eV for both $d_{xz}$ and $d_{yz}$ orbitals, and the NNN interorbital hopping is $\pm 0.07$ eV. $KS\simeq 0.4 - 0.8$ eV. $J S^2$ and $J' S^2$ are of the same order of $10$ meV. Since the Se anion is located above the center of the Fe plaquette, hybridizations via the Fe-As-Fe path give rise to comparable NN and NNN parameters.

To get a general and simple picture about the magnetic landscaping of the model, we compared a variety of static spin orders, such as the ferromagnetic (FM) state and the AF states of \textit{C}-type, \textit{E}-type, and \textit{G}-type (i.e., the N\'{e}el state where all NN spins are antiparallel), with the localized spins treated as Ising spins \cite{_DE:11:Yin,_strong:Hansmann}. Then, Eq. (1) is reduced to a system of noninteracting electrons moving in an external potential that is $-KS/2$ and $KS/2$ at site $i$
when the itinerant electron is spin parallel and antiparallel to $\vec {S}_i$, respectively.

The resulting phase diagram of the model is presented in Fig.~\ref{fig3}. Changing the spin moment $S$ will lead to changing both $KS$ and $J' S^2$. This direction of change, as marked by the red ray in Fig.~\ref{fig3}, will not induce the $C$-$E$ transition. On the other hand, we argued that changing $z_\mathrm{anion}$ corresponds to changing $KS$, as marked by the black ray in Fig.~\ref{fig3}: Since the iron atoms communicate with each other through the anions, the farther away the anions are, the more isolated the iron atoms are. The isolation of the Fe atoms would enhance the local parameter $S$ but suppress the nonlocal parameters $J_{ij}$; therefore, $J_{ij}S^2$ as a whole is much less material dependent than $KS$. Hence, the effective Hund's rule coupling term is decisive in determining the ground-state magnetic pattern. (Note that the enhancement of the on-site Coulomb and Hund's rule interaction terms by increasing $z_\mathrm{anion}$ were also found in the Hubbard model of the 10-fold Fe-3$d$ bands for FeSCs \cite{dft:model:Miyake}.) The results of the spin-fermion model agree well with those from first-principles presented in Fig.~\ref{fig2}.

The Hund's rule coupling brings in a blockade mechanism: The electron hopping to a neighboring site with opposite spin orientation suffers from the energy barrier whose height is of order of $KS$. Hence, the larger $KS$ is, the more likely the neighboring spins are aligned to the same direction. This is the so-called double-exchange ferromagnetism \cite{Mn:Zener,Mn:Anderson}, which has been studied intensively and extensively in the context of manganites for their colossal magnetoresistance phenomenon. In addition, comparable $J$ and $J'$ between the localized spins favor the $C$ type. Hence, FeSCs contain intrinsic competition between the double-exchange ferromagnetism and the $C$-type superexchange antiferromagnetism in the localized-spins sector. The $E$ type emerges in the intermediate regime as a compromise.

This model also reproduces the weak orbital polarization within the $xz$ and $yz$ (or $XZ$ and $YZ$) orbitals in the $E$ type because (i) the on-site occupation numbers are always the same in both the $xz$ and $yz$ orbitals in the $E$ type from symmetry consideration, and (ii) the parameters that can mix these two orbitals to polarize in the $XZ$ and $YZ$ representation are negligibly small.

\section{Discussion}

The double-exchange ferromagnetism was employed in the previous proposal that the metallic bicollinear antiferromagnetism in FeTe was driven by a site orbital ordering \cite{_oo:Turner:11}. However, the double-exchange effect was treated as being  secondary to the $YZ$ ferro-orbital order and the AF spin order along the $Y$ direction; it was used to introduce weak FM ordering of those AF $Y$ chains along the $X$ direction. This way actually decoupled the whole Fe planar lattice into two interpenetrating sublattices with each one exhibiting the $C$-type AF order on its own. This means that the NN exchange coupling strength is weak. On the contrary, in the spin-fermion model Eq. (\ref{eq1}) for the $E$-type FeTe $KS\simeq 0.8$ eV is the leading energy scale; to gain Hund's rule energy is so important that the NN Fe-Fe bonding along the zigzag FM chain (see Fig.~1) is strong. This means that if one has to fit the magnetic energy surface of FeTe to the Heisenberg model, the resulting NN exchange coupling on average is strong and FM, in agreement with neutron scatter data on FeTe \cite{neu:11:Lipscombe}.

The spin-fermion model Eq. (\ref{eq1}) appears in form similar to that used to describe the manganites. These two classes of materials do share some common features, such as magnetic softness and large normal-state resistivity. We emphasize two differences between FeSCs and manganites. Firstly, the Jahn-Teller distortion energy in FeSCs \cite{dft:1111:Lee} is one order of magnitude smaller than in manganites \cite{Mn:lamno3:Yin} because the layered structure of edge-sharing anion tetrahedra surrounding Fe atoms is much less flexible than the network of corner-sharing octahedra surrounding Mn atoms. As a result, unlike in the manganites where the orbital degree of freedom is often frozen into an orbital order via the cooperative Jahn-Teller effect, there lacks such a locking mechanism in FeSCs. A small change in $KS$ could lead to a dramatic change in the orbital order status and thus would substantially affect the fit to the Heisenberg model. In fact, neutron scattering data on CaFe$_2$As$_2$ \cite{neu:122:Zhao} and FeTe \cite{neu:11:Lipscombe} reported that the $C$ and $E$ types were surprisingly well separated in the Heisenberg model parameter space, with the leading NN exchange interaction being AF and FM, respectively. This implies that while the Heisenberg model is still reasonable for describing magnetic linear response near one particular ground state, it cannot capture the essential orbital physics and is problematic for describing the general magnetic softness in FeSCs. Secondly, $KS\simeq 2$ eV in manganites is much larger, driving the system into the FM regime, as shown in Fig.~3. With a moderate $KS$, FeSCs are located near the $C$-$E$ phase boundary and the FM phase becomes an irrelevant high-energy state, providing a necessary environment for the formation of singlet superconductivity where the paired electrons have opposite spins.

Regarding superconductivity, this magnetic softness is actually a two-blade sword. On the one hand,
with its strong ability to undertake electronic reconstruction, the system may on the one hand
manage to efficiently screen direct electron-electron Coulomb interaction and generate
appropriate bosonic excitations to glue electron pairs in order to form superconductivity. The magnetic softness also implies a large scattering rate for electron transport and the observed large resistivity in the normal state. This is a favorable feature, according to Homes's law that $T_c$ is proportional to both the superfluid density and the normal-state resistivity \cite{Homes:Nature}.
On the other hand, a too correlated system may find ways to satisfy the competing
non-superconducting players first. For example, the $C_xE_{1-x}$ region, a mixed $C$-type and
$E$-type AF, could be formed in accord with the change in doping level \cite{Mn:Hotta,neu:11:Xu}. Only until none of
these players can be adequately satisfied could superconductivity show up to relief the high entropy---likely with a low superfluid
density. While the high-$T_c$ mechanism of superconductivity in both FeSCs
and cuprates remains to be discovered, the present results suggest that FeSCs, though
closer kin to the manganites than the cuprates in terms of their magnetism, can exhibit a quantum phase transition to superconductivity instead of ferromagnetism.

\ignore{
\begin{figure}[t]
\center{
\includegraphics[width=0.9\columnwidth,clip=true,angle=0]{fig4.png}
}
\caption{\label{fig3} Electronic structures of the itinerant
electrons for the (a) collinear, (b) bicollinear, (c) checkerboard,
and (d) ferromagnetic spin orders calculated with KS ¼ 0:8 eV.
The dashed lines are the Fermi level for n ¼ 1. The energy unit
is eV.}
\end{figure}
}

\section{Summary}
We examine the relevance of the anion height from the iron plan, the ordered magnetic moment, and the orbital ordering pattern to the magnetic softness in iron-base superconductors by first-principles electronic structure analysis of their parent compounds. We conclude that the anion height is an effective tuning parameter and the others are not. These results are shown to be compatible with a recently proposed spin-fermion model where localized spins and orbitally degenerate itinerant electrons coexist and are coupled by Hund's rule coupling. This implies that the difference in the strength of the Hund's rule coupling term is the major material-dependent microscopic parameter for determining the ground-state spin pattern, and that the magnetic softness in iron-based superconductors is essentially driven by the competition between the double-exchange ferromagnetism and the superexchange antiferromagnetism.

\section{Acknowledgement}
This work was supported by the U.S. Department of Energy, Office of Basic Energy Science, under Contract No. DE-AC02-98CH10886.

\section*{References}


\end{document}